\documentclass[10pt]{article}
\usepackage{graphicx}
\usepackage{amsmath}
\usepackage{amssymb}
\usepackage{caption2}
\begin{document}
\bibliographystyle{prsty}
\begin{center}
{\large {\bf \sc{  Fully-heavy hexaquark  states via  the QCD sum rules }}} \\[2mm]
Zhi-Gang  Wang \footnote{E-mail: zgwang@aliyun.com.  }   \\
 Department of Physics, North China Electric Power University, Baoding 071003, P. R. China
\end{center}

\begin{abstract}
We construct the diquark-diquark-diquark type vector six-quark currents to investigate the vector and scalar hexaquark states  in the framework of the QCD sum rules in a consistent way, and make predictions for the hexaquark masses.    We can search for the fully-heavy hexaquark  states in the  di-$ \Omega_{ccc}$ and
 di-$ \Omega_{bbb}$  invariant mass spectra  in the future. Combined with our previous works on the fully-heavy tetraquark states and pentaquark states,
  the present predictions  can shed light on the nature of the fully-heavy multiquark states.

\end{abstract}



\section{Introduction}
In recent years, there has been  great progress on the spectroscopy of the exotic states, which cannot find their suitable positions in the conventional spectra of the two-quark and three-quark models \cite{PDG}. The exotic states provide us with  an excellent  subject to explore
the strong interactions  governing  dynamics of the quarks and gluons, and  confinement mechanism. It is interesting and necessary  to explore  the fully-heavy tetraquark (molecular) states, pentaquark (molecular) states and hexaquark (molecular) states,  in those cases,  we do not
have to deal with the complex dynamics involving both explicit  heavy and light degrees of freedoms.

Experimentally, in 2018, the LHCb collaboration  investigated the $\Upsilon\,\mu^+\mu^-$ invariant-mass distribution  for a possible exotic  state $X_{bb\bar{b}\bar{b}}$  based on a data sample of proton-proton collisions recorded with the LHCb detector at $\sqrt{s} = 7$, $8$ and $13\,\rm{TeV}$ corresponding to an integrated luminosity of
  ${\rm 6.3fb}^{-1}$, and observed  no significant excess of events  \cite{LHCb-bbbb-1806}.

In 2020, the CMS collaboration searched for narrow resonances decaying to the final states $\Upsilon\,\mu^+\mu^-$ using proton-proton collision data collected in 2016 by the CMS experiment corresponding to an integrated luminosity of $35.9 {\rm fb}^{-1}$, and observed no  significant excess of events \cite{CMS-bbbb-2002}.

Also, in 2020, the LHCb collaboration  observed a  narrow structure $X(6900)$  and
  a broad structure just above the di-$J/\psi$ threshold in the di-$J/\psi$   invariant mass distributions with the  statistical  significance  larger than $5\sigma$ using proton-proton collision data at center-of-mass energies of $\sqrt{s} = 7$, $8$ and $13\,\rm{TeV}$ corresponding to an integrated luminosity of
  ${\rm 9fb}^{-1} $\cite{LHCb-cccc-2006},  such resonance structures  are
 the first fully-heavy exotic multiquark candidates observed  experimentally up to today.

The observation of the $X(6900)$ sheds some  light on the nature of the  exotic states and stimulates  much interests  in (and many works on) the fully-heavy multiquark states, such as the tetraquark (molecular) states, pentaquark (molecular) states, hexaquark (molecular) states, etc. There are four charm valence quarks considering its  observation in the di-$J/\psi$ mass spectrum. The di-charmonium thresholds are $2M_{\eta_c}=5.968\,\rm{GeV}$, $2M_{J/\psi}=6.194\,\rm{GeV}$, $2M_{\chi_{c0}}=6.829\,\rm{GeV}$, $2M_{\chi_{c1}}=7.021$, $2M_{\chi_{c2}}=7.112\,\rm{GeV}$, $2M_{h_c}=7.051\,\rm{GeV}$ from the Particle Data Group \cite{PDG}, they lie either below or above the threshold of the $X(6900)$, it is difficult to assign the $X(6900)$ as the loosely bound tetraquark molecular state without introducing the (large) coupled-channel effects \cite{GuoFK-cccc,Couple-C-X6900-Guo,Couple-C-X6900-Zheng}, it is more natural to assign the $X(6900)$ as the excited diquark-antidiquark type tetraquark state \cite{WZG-di-di-CPC,Bedolla-di-di-cccc,LuQF-di-di-cccc,Lebed-di-di-cccc,Rosner-di-di-cccc,Zhu-di-di-cccc,X6900-di-di-PJL,X6900-di-di-PJL-2,X6900-di-di-Maiani,X6900-di-di-V,X6900-di-di-Chao,X6900-di-di-Faustov,
X6900-di-di-ZHURL}.

Usually, we expect that  the loosely bound tetraquark molecular states have the spatial extension larger than $1\,\rm{fm}$. While in the QCD sum rules, we choose the local four-quark currents, which couple potentially to the compact color-singlet-color-singlet type tetraquark states, rather than to the loosely bound tetraquark molecular states, although we also call them as tetraquark molecular states in the QCD sum rules \cite{WZG-local,WZG-Comment}.

  The attractive force  induced by  one-gluon exchange  favors  forming    the diquark correlations in  color antitriplet, Fermi-Dirac  statistics requires  $\varepsilon^{abc} Q^{T}_b C\Gamma Q_c
  =\varepsilon^{abc} Q^{T}_b \left[C\Gamma\right]^T Q_c$,  $\left[C\Gamma\right]^T=-C\Gamma$ for $\Gamma=\gamma_5$, $1$,
  $\gamma_\mu\gamma_5$;  $\left[C\Gamma\right]^T=C\Gamma$ for  $\Gamma=\gamma_\mu$, $\sigma_{\mu\nu}$, where the $a$, $b$ and $c$ are color indexes. Only the   axialvector  diquarks $\varepsilon^{abc} Q^{T}_b C\gamma_\mu Q_c$ and tensor diquarks  $\varepsilon^{abc} Q^{T}_b C\sigma_{\mu\nu }Q_c$ can exist.
   The diquark operators $\varepsilon^{abc}Q^{T}_bC\sigma_{\mu\nu} Q_c$ have both the spin-parity  $J^P=1^+$ and $1^-$ components, for the spin-parity $J^P=1^-$ component, there exists an implicit P-wave which is embodied in the negative parity, the axialvector diquark correlations   are more stable than the tensor diquark correlations due to the additional energy excited by the P-wave. To obtain the lowest masses, we usually  take the axialvector diquark operators $\varepsilon^{abc}  Q^T_bC\gamma_\mu Q_c$ as the elementary constituents  to explore the
 doubly-heavy (triply-heavy) baryon states, tetraquark states and pentaquark states
 \cite{WZG-AAPPS,WZG-QQQQ-EPJC-1,WZG-QQQQ-EPJC-2,WZG-QQ-EPJC-2}.

The tetraquark candidate $X(6900)$ was observed in the di-$J/\psi$ invariant mass spectrum \cite{LHCb-cccc-2006}, the pentaquark candidates $P_c(4312)$, $P_c(4380)$, $P_c(4440)$, $P_c(4457)$ were observed in the $J/\psi p$ invariant mass spectrum \cite{LHCb-4380,LHCb-Pc4312}.
Analogously, we expect to observe the fully-heavy pentaquark  candidates $QQQQ\bar{Q}$ in the $J/\psi \Omega_{ccc}$ and $\Upsilon \Omega_{bbb}$ invariant mass spectra, expect to observe the fully-heavy hexaquark and dibaryon   candidates $QQQQQQ$ in the $\Omega_{ccc} \Omega_{ccc}$ and $\Omega_{bbb} \Omega_{bbb}$ invariant mass spectra, expect to observe the fully-heavy baryonium candidates in the $\Omega_{ccc} \overline{\Omega}_{ccc}$ and $\Omega_{bbb} \overline{\Omega}_{bbb}$ invariant mass spectra at the LHCb,  CEPC, FCC and ILC in the future.

Theoretically, there have been several works on the fully-heavy exotic states with more than four valence quarks.
In Ref.\cite{ZhangJR-QQQQQ}, J. R. Zhang explores the color-singlet-color-singlet type $\eta_c \Omega_{ccc}$ and $\eta_b \Omega_{bbb}$ pentaquark molecular states with the QCD sum rules. In Ref.\cite{WZG-Q5-NPB}, we construct the diquark-diquark-antiquark type five-quark currents to study  the fully-heavy pentaquark states in the framework of the QCD sum rules. In Ref.\cite{An-D-QQQQQ},
H. T. An et al explore the fully-heavy pentaquark states  with the modified chromo-magnetic interaction model .

As far as the hexaquark states are  concerned, in Ref.\cite{Omega-Omega-BS-Latt}, Y. Lyu et al study the dibaryon  state $\Omega_{ccc}\Omega_{ccc}({}^1S_0)$ based on the (2+1)-flavor lattice QCD and observe that there exists a shallow dibaryon  state $\Omega_{ccc}\Omega_{ccc}({}^1S_0)$ with the binding energy about
$5.7\,\rm{MeV}$ and spatial extension about $\sqrt{\langle r^2\rangle}\approx 1.1\,\rm{fm}$. In Ref.\cite{Omega-Omega-GLS}, Liu and  Geng explore the $\Omega\Omega$, $\Omega_{ccc}\Omega_{ccc}$ and $\Omega_{bbb}\Omega_{bbb}$  dibaryon states in the extended one-boson exchange model. In Ref.\cite{Omega-Omega-PJL}, H. Huang et al
investigate the existence of the fully-heavy dibaryon states $\Omega_{ccc}\Omega_{bbb}$, $\Omega_{ccc}\Omega_{ccc}$ and $\Omega_{bbb}\Omega_{bbb}$ with the spins  $J=0$, $1$, $2$, $3$  using  a constituent quark model, and find that only the dibaryon state composed of six $c$ or $b$ quarks with the spin-parity $J^{P}=0^{+}$ can be bound. If there exist the $\Omega_{QQQ}\Omega_{QQQ}$ dibaryon states, naively, we expect that those color-singlet-color-singlet type hexaquark states lie below the di-$\Omega_{QQQ}$ thresholds, unfortunately, the triply-heavy baryon states  still escape the experimental detections up to today \cite{WZG-AAPPS}.

In  previous works, we constructed the color-singlet-color-singlet type (color-singlet-color-singlet-color-singlet type) six-quark  currents to explore the triply-heavy dibaryon (hexaquark molecular) states and doubly-heavy dibaryon states with the QCD sum rules \cite{WZG-ccc-dibaryion,WZG-cc-dibaryon,WZG-cc-dibaryon-2}  (\cite{WZG-DvDvDV}), and constructed  the diquark-diquark-diquark type six-quark currents to explore the triply-heavy hexaquark state with the QCD sum rules \cite{WZG-tri-ccc-hexa}. In all the works  \cite{WZG-ccc-dibaryion,WZG-cc-dibaryon,WZG-cc-dibaryon-2,WZG-DvDvDV,WZG-tri-ccc-hexa}, there are explicit light degrees of freedoms  in the six-quark currents,
 the contributions from the valence light quarks are very large.

 In this work, we construct the diquark-diquark-diquark type local six-quark currents to study  the fully-heavy hexaquark states with the spin-parity $J^P=1^-$ and $0^+$ consistently in the framework of the QCD sum rules, and make predictions for the hexaquark masses to be confronted to the experimental data in the future,
because  the QCD sum rules approach is a powerful theoretical tool in studying  the exotic $X$, $Y$, $Z$ and $P$ states \cite{MNielsen-review-1812}.

The article is arranged as follows:  in Sect.2, we acquire the QCD sum rules for   the  fully-heavy hexaquark states; in Sect.3, we present the numerical results and discussions; Sect.4 is reserved for our conclusion.

\section{QCD sum rules for  the  fully-heavy hexaquark states  }
Firstly, we  write down  the correlation functions  $\Pi_{\mu\nu}(p)$,
\begin{eqnarray}
\Pi_{\mu\nu}(p)&=&i\int d^4x e^{ip \cdot x} \langle0|T\left\{J_{\mu}(x)J_{\nu}^{\dagger}(0)\right\}|0\rangle \, ,
\end{eqnarray}
where
\begin{eqnarray}
J_\mu(x)&=&\varepsilon^{ijk}\varepsilon_{\mu\nu\alpha\beta}\, A^\nu_i(x)\,A_j^{\alpha}(x)\,A^\beta_k(x)\, , \nonumber\\
A_i^\mu(x)&=& \varepsilon^{ijk}Q^{T}_j(x)C\gamma^\mu Q_k(x)\, ,
\end{eqnarray}
$Q=b$, $c$,  the $i$, $j$, $k$ are color indexes. We can construct other  diquark-diquark-diquark type six-quark currents consisting  of the same flavor
to interpolate the fully-heavy hexaquark states,
\begin{eqnarray}
\eta_\mu{}^{\nu^\prime\alpha^\prime\beta^\prime}(x)&=&\varepsilon^{ijk}\varepsilon_{\mu\nu\alpha\beta}\, T^{\nu\nu^{\prime}}_i(x)\,T_j^{\alpha\alpha^{\prime}}(x)\,T^{\beta\beta^\prime}_k(x)+ \cdots  \, , \nonumber\\
T_i^{\mu\nu}(x)&=& \varepsilon^{ijk}Q^{T}_j(x)C\sigma^{\mu\nu} Q_k(x)\, ,
\end{eqnarray}
where the Lorentz indexes $\nu^\prime$, $\alpha^\prime$, $\beta^\prime$ are symmetric.
The tensor diquark operators $T_i^{\mu\nu}(x)$ have both the spin-parity  $J^P=1^+$ and $1^-$ components, the P-wave effect is embodied in the negative parity, the additional P-wave leads  to less stable tensor diquark correlations compared to the  axialvector diquark correlations.  Therefore, the currents $\eta_\mu{}^{\nu^\prime\alpha^\prime\beta^\prime}(x)$ couple potentially to the fully-heavy hexaquark states with much (or slightly) larger masses than that of the currents $J_\mu(x)$.

 Under parity transform $\widehat{P}$, the currents $J_\mu(x)$ have the  property,
\begin{eqnarray}
\widehat{P} J_i(x)\widehat{P}^{-1}&=& - J_i(\tilde{x}) \, , \nonumber\\
\widehat{P} J_0(x)\widehat{P}^{-1}&=& + J_0(\tilde{x}) \, ,
\end{eqnarray}
where  $x^\mu=(t,\vec{x})$ and $\tilde{x}^\mu=(t,-\vec{x})$.
 The currents $J_{\mu}(x)$  couple potentially to both the vector and scalar fully-heavy hexaquark states, the correlation functions $\Pi_{\mu\nu}(p)$ at the hadron side can be written as,
 \begin{eqnarray}
 \Pi_{\mu\nu}(p)&=&\frac{\lambda_V^2}{M_V^2-p^2}\left(-g_{\mu\nu}+\frac{p_{\mu}p_{\nu}}{p^2} \right)+\frac{\lambda_S^2}{M_S^2-p^2}\frac{p_{\mu}p_{\nu}}{p^2}+\cdots \, ,\nonumber\\
 &=&\Pi_V(p^2)\left(-g_{\mu\nu}+\frac{p_{\mu}p_{\nu}}{p^2} \right)+\Pi_S(p^2)\frac{p_{\mu}p_{\nu}}{p^2}\, ,
 \end{eqnarray}
 where the pole residues $\lambda_V$ and $\lambda_S$ are defined by,
\begin{eqnarray}
\langle 0| J_\mu (0)|V(p)\rangle &=&\lambda_{V}\, \varepsilon_\mu \, , \nonumber \\
\langle 0| J_\mu (0)|S(p)\rangle &=&\lambda_{S}\, \frac{p_\mu}{\sqrt{p^2}} \, ,
\end{eqnarray}
 the $\varepsilon_\mu$ are the  polarization vectors of the vector fully-heavy hexaquark states.

\begin{figure}
 \centering
 \includegraphics[totalheight=5.0cm,width=12cm]{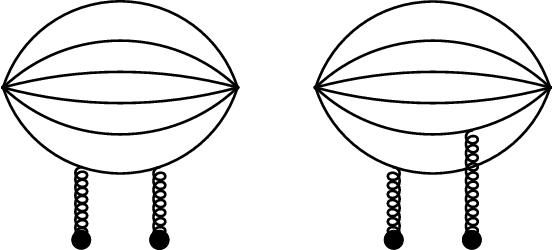}
      \caption{The Feynman diagrams make contributions   to the gluon condensates. Other
Feynman diagrams obtained by interchanging of the $Q$ quark lines  are implied. }\label{Feynman}
\end{figure}

We accomplish the tedious and terrible   operator product expansion and take account of the gluon condensates, then obtain the spectral densities $\frac{{\rm Im \, \Pi}_{V/S}(s)}{\pi}$ at the quark level through dispersion relation, \begin{eqnarray}
\Pi_{V/S}(p^2)&=&\frac{1}{\pi}\int_{36m_Q^2}^{\infty} ds \frac{{\rm Im \, \Pi}_{V/S}(s)}{s-p^2}\, .
\end{eqnarray}
We match the hadron side with the QCD side of the correlation functions $\Pi_{V/S}(p^2)$ below the continuum thresholds $s_0$ and perform the Borel transform with respect to the $P^2=-p^2$  to  obtain  the  QCD sum rules:
\begin{eqnarray}\label{QCDSR}
\lambda_{V/S}^2\exp\left( -\frac{M_{V/S}^2}{T^2}\right)&=& \int_{36m_Q^2}^{s_0}ds \int_{16m_Q^2}^{(\sqrt{s}-2m_Q)^2}dr \int_{4m_Q^2}^{(\sqrt{r}-2m_Q)^2}dt_1
\int_{4m_Q^2}^{(\sqrt{r}-\sqrt{t_1})^2}dt_2\int_{4m_Q^2}^{(\sqrt{s}-\sqrt{r})^2}dt_3\nonumber\\
&&\rho_{V/S}(s,r,t_1,t_2,t_3)\exp\left( -\frac{s}{T^2}\right)\,  ,
\end{eqnarray}
where
\begin{eqnarray}\label{QCD-density}
\rho_{V/S}&=& \frac{\sqrt{\lambda(s,r,t_3)}}{s}\frac{\sqrt{\lambda(r,t_1,t_2)}}{r} \frac{\sqrt{\lambda(t_1,m_Q^2,m_Q^2)}}{t_1}
\frac{\sqrt{\lambda(t_2,m_Q^2,m_Q^2)}}{t_2}\frac{\sqrt{\lambda(t_3,m_Q^2,m_Q^2)}}{t_3}\nonumber\\
&&\left(  C_{V/S,4}\, m_Q^{4}+ C_{V/S,2} \, m_Q^{2}+ C_{V/S,0}\right)\nonumber\\
&&+\langle\frac{\alpha_sGG}{\pi}\rangle \frac{\sqrt{\lambda(s,r,t_3)}}{s}\frac{\sqrt{\lambda(r,t_1,t_2)}}{r} \frac{\sqrt{\lambda(t_1,m_Q^2,m_Q^2)}}{t_1}
\frac{\sqrt{\lambda(t_2,m_Q^2,m_Q^2)}}{t_2}\nonumber\\
&&\frac{1}{t_3\sqrt{\lambda(t_3,m_Q^2,m_Q^2)}^5}\left(  C^{gg}_{V/S,10}\, m_Q^{10}+ C_{V/S,8}^{gg} \, m_Q^{8}+ C_{V/S,6}^{gg}\, m_Q^{6}+ C_{V/S,4}^{gg} \, m_Q^{4}+ C_{V/S,2}^{gg}m_Q^2\right)\nonumber\\
&&+\langle\frac{\alpha_sGG}{\pi}\rangle \frac{\sqrt{\lambda(s,r,t_3)}}{s}\frac{\sqrt{\lambda(r,t_1,t_2)}}{r} \frac{\sqrt{\lambda(t_1,m_Q^2,m_Q^2)}}{t_1}
\frac{\sqrt{\lambda(t_2,m_Q^2,m_Q^2)}}{t_2}\nonumber\\
&&\frac{1}{t_3\sqrt{\lambda(t_3,m_Q^2,m_Q^2)}^3}\left(  C_{V/S,8}^{g-g} \, m_Q^{8}+ C_{V/S,6}^{g-g}\, m_Q^{6}+ C_{V/S,4}^{g-g} \, m_Q^{4}+ C_{V/S,2}^{g-g}m_Q^2\right)\, ,
\end{eqnarray}
$\lambda(a,b,c)=a^2+b^2+c^2-2ab-2bc-2ac$, the $T^2$ is the Borel parameter, the lengthy/cumbersome  expressions of the coefficients $C_{V/S,4}$, $C_{V/S,2}$, $C_{V/S,0}$,  $C^{gg}_{V/S,10}$, $ C_{V/S,8}^{gg}$,
 $C_{V/S,6}^{gg}$, $C_{V/S,4}^{gg}$, $C_{V/S,2}^{gg}$, $C_{V/S,8}^{g-g}$, $C_{V/S,6}^{g-g}$, $C_{V/S,4}^{g-g}$, $C_{V/S,2}^{g-g}$ are neglected for simplicity, the interested readers    can get them in the Fortran form   through contacting  me via E-mail. In the following, we give an example to illustrate the function forms of the coefficients,
 \begin{eqnarray}\label{Coeff}
 C_{S,0}&=&\frac{3}{512\pi^{10}}\left[t_2t_3(t_2+t_3)-t_1t_3(t_1+t_3)-9t_1t_2(r+t_3)+\frac{t_3}{2}(s+r)(t_1-t_2)  +\frac{9st_1t_2}{2}\right. \nonumber\\
 &&-\frac{st_3}{r}(t_1^2-t_2^2)+\frac{2t_3^2}{r}(t_1^2-t_2^2)-\frac{9rt_1t_2t_3}{s}+\frac{rt_2t_3}{s}(t_2+t_3)
 +\frac{2t_3^2}{s}(t_1^2-t_2^2)-\frac{rt_1t_3}{s}(t_1+t_3)\nonumber\\
 &&-\frac{3t_1t_2t_3}{2r}(t_1-t_2)+\frac{t_3}{2r}(t_1^3-t_2^3)+\frac{9t_1t_2}{2s}(r^2+t_3^2)-\frac{3t_1t_2t_3}{2s}(t_1-t_2)
 +\frac{r^2t_3}{2s}(t_1-t_2)\nonumber\\
 &&+\frac{t_1t_3}{2s}(t_1^2+t_3^2)-\frac{t_2t_3}{2s}(t_2^2+t_3^2)+\frac{3t_1t_2t_3^2}{rs}(t_1-t_2)
 +\frac{t_2^2t_3^2}{sr}(t_2+t_3)-\frac{t_1^2t_3^2}{rs}(t_1+t_3)\nonumber\\
 &&+\frac{3t_1t_2t_3^2}{r^2}(t_1-t_2)-\frac{t_1^2t_3^2}{r^2}(t_1-t_3)-\frac{3st_1t_2t_3}{2r^2}(t_1-t_2)
 +\frac{st_3}{2r^2}(t_1^3-t_2^3)+\frac{t_3^3}{2r^2s}(t_1^3-t_2^3)\nonumber\\
 &&\left.-\frac{3t_1t_2t_3^3}{2r^2s}(t_1-t_2)  \right]\, .
 \end{eqnarray}
 It is the most simple coefficient, the total expressions of all the coefficients will cost  more pages than the whole text.
 The superscripts $gg$ and $g-g$ in the coefficients $C^{gg}_{V/S,10}$, $ C_{V/S,8}^{gg}$,
 $\cdots$,   $C_{V/S,8}^{g-g}$, $C_{V/S,6}^{g-g}$, $\cdots$ correspond to the Feynman diagrams where the two gluons
 which form the gluon condensates  are emitted from one heavy-quark line
 (the first Feynman diagram in Fig.\ref{Feynman})
 and two heavy-quark lines  (the second Feynman diagram in Fig.\ref{Feynman}),
 respectively. In calculations, we expand the QCD spectral densities $\rho_{V/S}(s,r,t_1,t_2,t_3)$ in terms of the powers of the $m_Q^2$, such as $m_Q^0$, $m_Q^2$, $m_Q^4$, $m_Q^6$, $\cdots$, the coefficients $C_{V/S,4}$, $C_{V/S,2}$, $C_{V/S,0}$,  $C^{gg}_{V/S,10}$, $\cdots$ are functions of the variables $s$, $r$, $t_1$, $t_2$ and $t_3$, just like that shown in Eq.\eqref{Coeff}. There are end-point divergences
 $\frac{1}{\sqrt{t_3-4m_Q^2}^5}$ and $\frac{1}{\sqrt{t_3-4m_Q^2}^3}$ because  $\lambda(t_3,m_Q^2,m_Q^2)=t_3\left(t_3-4m_Q^2 \right)$, the end-point divergence $\frac{1}{\sqrt{t_3-4m_Q^2}^5}$ is much worse than the end-point divergence   $\frac{1}{\sqrt{t_3-4m_Q^2}^3}$,
 we regulate the divergences by adding a uniform  mass term  $\Delta^{2}$ in the divergent terms  $\frac{1}{\sqrt{t_3-4m_Q^2+\Delta^2}^5}$ and $\frac{1}{\sqrt{t_3-4m_Q^2+\Delta^2}^3}$ with the value
 $\Delta^{ 2}=m_Q^2$ \cite{WZG-Q5-NPB}. In the Appendix, we  illustrate how to calculate the Feynman diagrams and illustrate the origins   of the end-point divergences.
  The three-gluon condensate makes tiny contribution in the Borel windows for the triply-heavy
 baryon states \cite{WZG-AAPPS}, such a tendency  survives for the fully-heavy exotic multiquark states.
 We neglect the three-gluon condensate, as it is the vacuum expectation value of the gluon operators of the order $\mathcal{O}(\alpha_s^{k})$ with $k=\frac{3}{2}$
 \cite{WZG-Hidden-charm-bottom-1,WZG-Hidden-charm-bottom-2,WZG-IJMPA-penta-1,WZG-IJMPA-penta-2}. Direct calculations indicate that the vacuum condensates which are  vacuum expectation values of the quark-gluon operators of the order $\mathcal{O}(\alpha_s^{k})$ with $\frac{3}{2}\leq k \leq 3$ can be neglected safely \cite{WZG-cc-dibaryon}.

We  differentiate  Eq.\eqref{QCDSR} in regard  to  $\frac{1}{T^2}$, then eliminate the
 pole residues $\lambda_{V/S}$ and obtain the  masses for the vector and scalar fully-heavy  hexaquark states,
 \begin{eqnarray}
 M^2_{V/S} &=&-\frac{\frac{d}{d(1/T^2)} \int_{36m_Q^2}^{s_0}ds \, \rho_{V/S;QCD}(s)\exp\left( -\frac{s}{T^2}\right)}{\int_{36m_Q^2}^{s_0}ds \, \rho_{V/S;QCD}(s)\exp\left( -\frac{s}{T^2}\right)}\, ,
\end{eqnarray}
where the $ \rho_{V/S;QCD}(s)$ are the corresponding spectral densities.

\section{Numerical results and discussions}
We choose  the  gluon condensate $\langle \frac{\alpha_s
GG}{\pi}\rangle=0.012\pm0.004\,\rm{GeV}^4$
\cite{SVZ79,PRT85,ColangeloReview}, and  take the $\overline{MS}$ masses of the heavy  quarks
$m_{c}(m_c)=(1.275\pm0.025)\,\rm{GeV}$ and $m_{b}(m_b)=(4.18\pm0.03)\,\rm{GeV}$
 from the Particle Data Group \cite{PDG}.
All the nonperturbative dynamics and confinement effects  are embodied in the running heavy quark masses and vacuum  condensates,
  we take  account of the energy-scale dependence of the $\overline{MS}$ masses,
 \begin{eqnarray}
 m_Q(\mu)&=&m_Q(m_Q)\left[\frac{\alpha_{s}(\mu)}{\alpha_{s}(m_Q)}\right]^{\frac{12}{33-2n_f}} \, ,\nonumber\\
\alpha_s(\mu)&=&\frac{1}{b_0t}\left[1-\frac{b_1}{b_0^2}\frac{\log t}{t} +\frac{b_1^2(\log^2{t}-\log{t}-1)+b_0b_2}{b_0^4t^2}\right]\, ,
\end{eqnarray}
  where $t=\log \frac{\mu^2}{\Lambda^2}$, $b_0=\frac{33-2n_f}{12\pi}$, $b_1=\frac{153-19n_f}{24\pi^2}$, $b_2=\frac{2857-\frac{5033}{9}n_f+\frac{325}{27}n_f^2}{128\pi^3}$,  $\Lambda=213\,\rm{MeV}$, $296\,\rm{MeV}$  and  $339\,\rm{MeV}$ for the quark flavor numbers  $n_f=5$, $4$ and $3$, respectively  \cite{PDG}.
We choose the flavors  $n_f=4$ and $5$ for the fully-heavy hexaquark states $cccccc$ and $bbbbbb$, respectively, and then
evolve the heavy quark masses  to the  typical energy scales $\mu=m_c(m_c)$ (i.e. $1.275\,\rm{GeV}$) and $2.8\,\rm{GeV}$ to extract the   masses of the
   hexaquark states $cccccc$ and $bbbbbb$, respectively, just like in our previous work on  the fully-heavy pentaquark states \cite{WZG-Q5-NPB}, and add an uncertainty  $\delta \mu=\pm 0.1\,\rm{GeV}$ \cite{WZG-Q5-NPB,WZG-DvDvDV}. Furthermore, we present the predictions  based on the updated gluon condensate  obtained by S. Narison,   $\langle \frac{\alpha_s
GG}{\pi}\rangle=0.021\pm0.001\,\rm{GeV}^4$ \cite{Narison-2101}.

We should choose suitable continuum thresholds $s_0$ to  avoid  contaminations from the first radial excited states and continuum states.  In previous
works, we choose $\sqrt{s_0}= M_{B}+0.50\sim 0.55\pm0.10\,\rm{GeV}$  for the triply-heavy baryon states $B$ \cite{WZG-AAPPS}, $\sqrt{s_0}=
M_{X}+0.50\pm0.10\,\rm{GeV}$  for the fully-heavy tetraquark states $X$
\cite{WZG-QQQQ-EPJC-1,WZG-QQQQ-EPJC-2}, $\sqrt{s_0}=M_{X/Z}+0.55\pm0.10\,\rm{GeV}$  for the
hidden-charm and hidden-bottom tetraquark states $X_Q$ and $Z_Q$ \cite{WZG-Hidden-charm-bottom-1,WZG-Hidden-charm-bottom-2},
$\sqrt{s_0}= M_{P}+0.65\pm0.10\,\rm{GeV}$  for the hidden-charm pentaquark states $P_c$ and $P_{cs}$
\cite{WZG-IJMPA-penta-1,WZG-IJMPA-penta-2},
$\sqrt{s_0}= M_{P}+0.60\pm0.10\,\rm{GeV}$  for the fully-heavy pentaquark states $P_Q$  \cite{WZG-Q5-NPB}. To obtain the lowest hexaquark masses, we prefer the smaller continuum threshold parameters and  require  the
  continuum threshold parameters satisfy the relation $\sqrt{s_0}= M_{V/S}+0.50\pm 0.10\,\rm{GeV}$ tentatively, and change
 the  Borel parameters and continuum threshold parameters  to acquire the best values  via trial and error.
We tentatively choose  a continuum threshold parameter $s_0$, then obtain the numerical value of the hexaquark mass $M_{V/S}$ from  the QCD sum rules, and  judge whether or not the two basic criteria of the QCD sum rules (plus the constraint  $\sqrt{s_0}=M_{V/S}+0.5\pm 0.1\,\rm{GeV}$) are satisfied. If not, we choose  another continuum threshold parameter $s_0$ until reach the satisfactory results.

At last,  we  obtain the best continuum threshold parameters, Borel parameters, and pole contributions, which are shown explicitly in Table \ref{BorelP}.
   In the Borel windows, the pole contributions are about $(40-60)\%$,  the pole dominance criterion is satisfied. On the other hand, the main contributions come from the perturbative terms,  the gluon condensate contributions play a minor role, and change slowly with variations of the Borel parameters, the convergent behaviors of the operator product expansion  are very good.
   In the Borel windows, the contributions of the gluon condensate are about $-2\%$ ($-4\%$) and $-1\%$ ($-2\%$) for the $J^P=1^-$ and $0^+$ fully-charm hexaquark states, respectively, and  $\ll 1\%$ for  the $J^P=1^-$ and $0^+$ fully-bottom hexaquark states, where the values in the brackets  come from the updated gluon condensate \cite{Narison-2101}. We choose the Borel windows with the help of the uniform pole contributions $(40-60)\%$, as the convergence of the operator product expansion is almost warranted  automatically.

\begin{figure}
 \centering
 \includegraphics[totalheight=6cm,width=7cm]{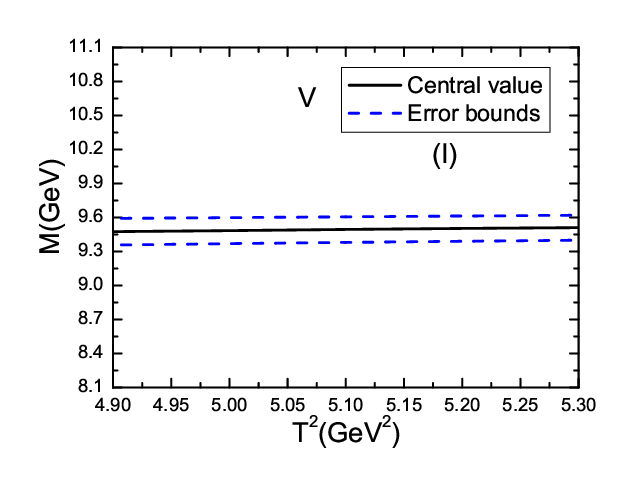}
\includegraphics[totalheight=6cm,width=7cm]{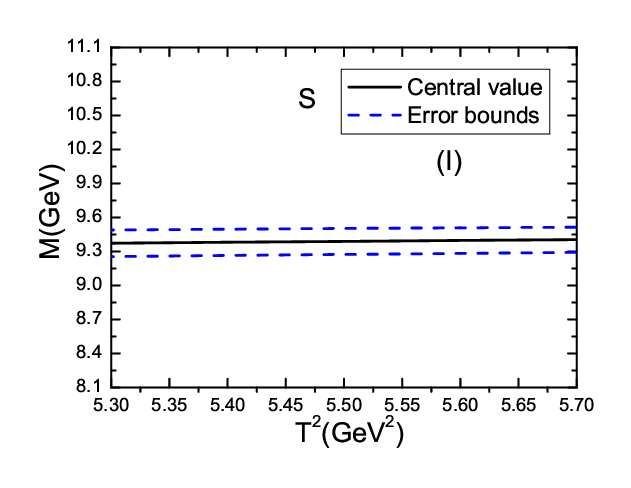}
\includegraphics[totalheight=6cm,width=7cm]{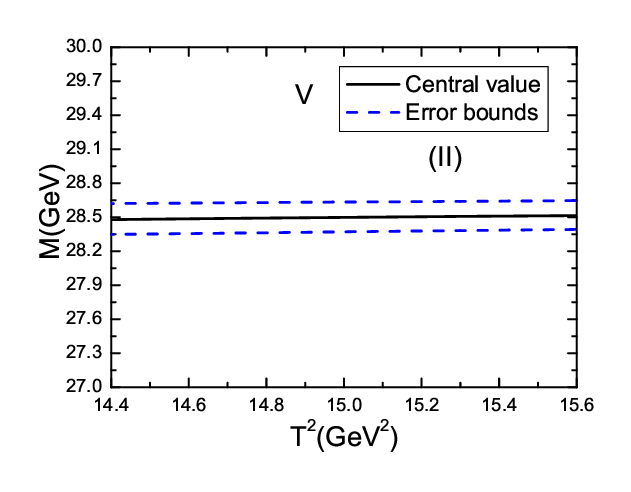}
\includegraphics[totalheight=6cm,width=7cm]{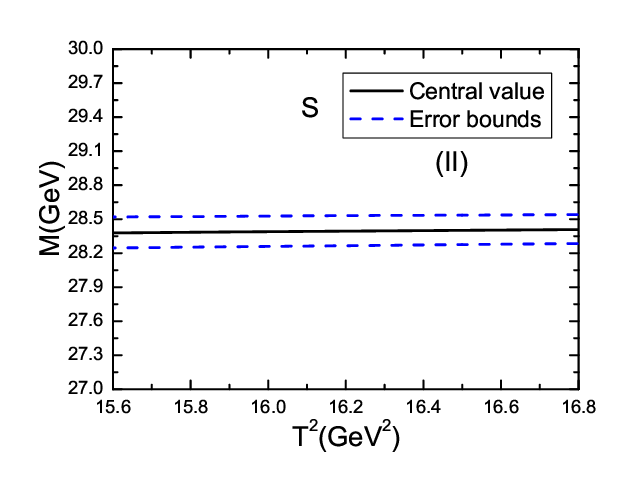}
 \caption{ The masses of the vector ($V$) and scalar ($S$) hexaquark states  with variations  of the Borel parameters $T^2$, where the (I) and (II) denote the
 $cccccc$ and $bbbbbb$ hexaquark states, respectively.  }\label{mass-cc-bb}
\end{figure}

     We  take account of all uncertainties of the relevant  parameters,  and acquire  the masses and pole residues of the
fully-heavy hexaquark  states, which are also shown plainly in Table \ref{BorelP} and Fig.\ref{mass-cc-bb}. In Table \ref{BorelP}, we also present the values of the masses and pole residues acquired  with the updated gluon condensate obtained by S. Narison \cite{Narison-2101}. From the Table, we can see clearly that the standard values and updated values lead to in-distinguishable  predictions, which emerge as a result of the tiny contributions of the gluon condensate.
 From Fig.\ref{mass-cc-bb}, we can see that there appear very flat platforms for the hexaquark masses, the uncertainties  come
  from the Borel parameters  are rather small, the predictions are robust.

In calculations,  we observe that the masses and pole residues increase monotonously and slowly with the increase of the continuum threshold parameters, we determine the continuum threshold parameters $s_0$ by adopting the  uniform constraints, such as the continuum thresholds  $\sqrt{s_0}= M_{V/S}+0.50 \pm0.1 \,\rm{GeV}$, pole contributions $(40\sim 60)\%$ and  intervals   $T^2_{max}-T^2_{min}=0.4\,\rm{GeV}^2$ ($1.2\,\rm{GeV}^2$) to acquire reliable predictions for the hexaquark states $cccccc$ ($bbbbbb$), where the  $T^2_{max}$ and $T^2_{min}$ stand for the maximum and minimum values of the Borel platforms, respectively.

\begin{table}
\begin{center}
\begin{tabular}{|c|c|c|c|c|c|c|c|c|}\hline\hline
              & $J^{P}$    & $T^2 (\rm{GeV}^2)$ & $\sqrt{s_0}(\rm GeV) $    &pole         &$M(\rm GeV)$    &$\lambda(\rm GeV^8)$ \\ \hline

$cccccc$      & $1^{-}$    & $4.9-5.3$          & $10.00\pm0.10$            &$(41-61)\%$  &$9.49\pm0.13$   &$(1.03\pm0.43)\times 10^{-1}$ \\
              &            &                    &                           &             &$9.50\pm0.13$                                                                       &$(1.03\pm0.43)\times 10^{-1}$ \\ \hline

$cccccc$      & $0^{+}$    & $5.3-5.7$          & $9.90\pm0.10$             &$(41-59)\%$  &$9.39\pm0.13$   &$(3.37\pm1.21)\times 10^{-1}$ \\

              &            &                    &                           &             &$9.39\pm0.13$   &$(3.36\pm1.21)\times 10^{-1}$ \\ \hline

$bbbbbb$      & $1^{-}$    & $14.4-15.6$        & $29.00\pm0.10$            &$(40-60)\%$  &$28.50\pm0.15$  &$5.61\pm2.70$ \\
              &            &                    &                           &             &$28.50\pm0.15$  &$5.61\pm2.70$ \\\hline

$bbbbbb$      & $0^{+}$    & $15.6-16.8$        & $28.90\pm0.10$            &$(41-59)\%$  &$28.39\pm0.15$  &$35.3\pm 15.6$ \\

              &            &                    &                           &             &$28.39\pm0.15$  &$35.3\pm 15.6$ \\ \hline

\hline\hline
\end{tabular}
\end{center}
\caption{ The Borel parameters, continuum threshold parameters, pole contributions, masses and pole residues for  the  ground state fully-heavy hexaquark states, the values in the lower lines are obtained from the updated gluon condensate. }\label{BorelP}
\end{table}

The thresholds of the tri-$J/\psi$, tri-$\eta_c$, tri-$\Upsilon$ and tri-$\eta_b$ are $3M_{J/\psi}=9.291\,\rm{GeV}$, $3M_{\eta_c}=8.952\,\rm{GeV}$, $3M_{\Upsilon}=28.381\,\rm{GeV}$ and $3M_{\eta_b}=28.196\,\rm{GeV}$ respectively from the Particle Data Group  \cite{PDG}, the central values have the relations,
\begin{eqnarray}
M_{cccccc}(1^-) &> &M_{cccccc}(0^+)> 3M_{J/\psi} >3M_{\eta_c} \, , \nonumber\\
M_{bbbbbb}(1^-) &> &M_{bbbbbb}(0^+)> 3M_{\Upsilon} >3M_{\eta_b} \, .
\end{eqnarray}
The fully-heavy hexaquark states $QQQQQQ$ have net positive or negative electric charges, naively, we expect that the repulsive electromagnetic interactions between the heavy quarks $Q$ and $Q$ lead to larger spatial extended bound states that of the heavy quark pairs $Q$ and $\bar{Q}$, where the attractive   electromagnetic interactions always favor forming tighter bound states, it is natural that the ground state masses have the relations  $M_{cccccc}(0^+)\geq3M_{J/\psi}$ and $M_{bbbbbb}(0^+)\geq3M_{\Upsilon}$. There exist an additional P-wave in the vector fully-heavy hexaquark states, which can excite some additional  energies,  so the masses have the hierarchies $M_{cccccc}(1^-) \geq M_{cccccc}(0^+)$ and $M_{bbbbbb}(1^-) \geq M_{bbbbbb}(0^+)$, the predictions are reasonable.

The decays of the fully-heavy hexaquark states can take place through the Okubo-Zweig-Iizuka super-allowed fall-apart mechanism,
\begin{eqnarray}
cccccc\, (1^-,0^+)&\to & \Omega_{ccc} \,\Omega_{ccc} \, , \nonumber\\
bbbbbb\, (1^-,0^+)&\to & \Omega_{bbb} \,\Omega_{bbb} \, ,
\end{eqnarray}
if there are enough phase-spaces.
According to the recent analysis of the  QCD sum rules
\cite{WZG-AAPPS}, $M_{\Omega(ccc)}=4.81 \pm 0.10\,\rm{GeV}$ and $M_{\Omega(bbb)}=14.43 \pm 0.09\,\rm{GeV}$, we  obtain the relations,
 \begin{eqnarray}
M_{cccccc}(0^+) &\leq &M_{cccccc}(1^-)\leq  2M_{\Omega(ccc)}=9.62 \pm 0.20\,\rm{GeV} \, , \nonumber\\
M_{bbbbbb}(0^+) &\leq &M_{bbbbbb}(1^-)\leq  2M_{\Omega(bbb)}=28.86 \pm 0.18\,\rm{GeV} \, ,
\end{eqnarray}
there only exist rather small phase-spaces due to uncertainties, the decays are  kinematically suppressed. In other words, the decays  can take place through the intermediate virtual $\Omega_{ccc}^*$ and $\Omega_{bbb}^*$ states,
\begin{eqnarray}
cccccc\, (1^-,0^+)&\to & \Omega_{ccc} \,\Omega_{ccc}^* \, , \nonumber\\
bbbbbb\, (1^-,0^+)&\to & \Omega_{bbb} \,\Omega_{bbb}^* \, .
\end{eqnarray}
In Ref.\cite{Omega-Omega-BS-Latt}, the studies based on the (2+1)-flavor lattice QCD indicate that the dibaryon  state $\Omega_{ccc}\Omega_{ccc}({}^1S_0)$ has the binding energy about $5.7\,\rm{MeV}$, i.e. it lies about $5.7\,\rm{MeV}$ below the di-$\Omega_{ccc}$ threshold, which is consistent with the present calculations, as a hadron maybe have several Fock components, we can choose several currents to interpolate it.
Experimentally, we can search for the fully-heavy hexaquark states in the di-$\Omega_{ccc}$ and di-$\Omega_{bbb}$
 invariant mass spectra  at the LHCb,   CEPC, FCC and ILC in the future.

 Unfortunately, up to now,  the  triply-heavy baryon states $\Omega_{ccc}$ and $\Omega_{bbb}$ still escape from experimental detections,
 and we can search for them in the decay chains, $\Omega_{ccc}\to \Omega_{ccs}\,\pi^+ \to \Omega_{css}\,\pi^+\pi^+\to \Omega_{sss}\,\pi^+\pi^+\pi^+$ and
 $\Omega_{bbb}\to \Omega_{bbs}\,J/\psi\to\Omega_{bss}\,J/\psi J/\psi\to\Omega_{sss}\,J/\psi J/\psi J/\psi$ through the weak decays $c \to s u\bar{d}$
and $b\to c\bar{c}s$ at the quark level. We  can search for the doubly/singly-heavy baryon states  $\Omega_{ccs}({\frac{1}{2}}^+)$,
$\Omega_{ccs}({\frac{3}{2}}^+)$, $\Omega_{bbs}({\frac{1}{2}}^+)$,
$\Omega_{bbs}({\frac{3}{2}}^+)$ and $\Omega_{bss}({\frac{3}{2}}^+)$ as a byproduct and additional benefit, as they have not been observed yet.

\section{Conclusion}
In this article,  we extend our previous works on the fully-heavy baryon states, tetraquark states and pentaquark states to explore the fully-heavy hexaquark states, and construct the diquark-diquark-diquark type vector local six-quark  currents to study the
vector and scalar hexaquark states consistently in the framework of the QCD sum rules. After accomplishing  the tedious  calculations,
 we obtain the  masses and pole residues of the fully-heavy hexaquark states.
 The central values of the predicted hexaquark masses
 $M_{cccccc}(1^-)=9.49\pm0.13\,\rm{GeV}$,
$M_{cccccc}(0^+)=9.39\pm0.13\,\rm{GeV}$,
$M_{bbbbbb}(1^-)=28.50\pm0.15\,\rm{GeV}$,
$M_{bbbbbb}(0^+)=28.39\pm0.15\,\rm{GeV}$ lies slightly below the corresponding di-$\Omega_{ccc}$ and di-$\Omega_{bbb}$ thresholds, respectively,
if we choose the masses of the
 $\Omega_{QQQ}$ states from the recent analysis of the QCD sum rules as the input parameters, the strong decays to the di-$\Omega_{ccc}$ and di-$\Omega_{bbb}$ induced states can take place through the intermediate virtual $\Omega_{ccc}^*$ and $\Omega_{bbb}^*$ states. We should bear in mind that the decays to the di-$\Omega_{QQQ}$ states are not forbidden due to the uncertainties.
   We can search for the fully-heavy hexaquark  states in the  di-$ \Omega_{ccc}$ and
 di-$ \Omega_{bbb}$  invariant mass spectra  at the  LHCb,  CEPC, FCC and ILC in the future, and confront the predictions to the experimental data.

\section*{Appendix}
Now we illustrate how to accomplish the operator product expansion, at the lowest order,
\begin{eqnarray}
\Pi_{\mu\nu}(p)&=& \int d^4k_1 d^4k_2 d^4k_3 d^4k_4 d^4k_5 d^4k_6\,\delta^4\left(p-k_1-k_2-k_3-k_4-k_5-k_6 \right)\nonumber\\
&& \frac{ f_{\mu\nu}(k_1,k_2,k_3,k_4,k_5,k_6)}{\left(k_1^2-m_Q^2 \right)\left(k_2^2-m_Q^2 \right)\left(k_3^2-m_Q^2 \right)\left(k_4^2-m_Q^2 \right)\left(k_5^2-m_Q^2 \right)\left(k_6^2-m_Q^2 \right)} \, , \nonumber\\
&=&\int d^4q_1 d^4q_2 d^4q_3  \,\delta^4\left(p-q_1-q_2-q_3 \right) \nonumber\\
&& \frac{(-2\pi i)^2}{2\pi i} \int_{4m_Q^2}^{t_1^f} d t_1 \frac{1}{t_1-q_1^2}\int d^4k_1 d^4k_2\,\delta^4 \left(q_1-k_1-k_2 \right)\delta \left(k_1^2-m_Q^2 \right)\delta \left(k_2^2-m_Q^2 \right)  \nonumber \\
&& \frac{(-2\pi i)^2}{2\pi i} \int_{4m_Q^2}^{t_2^f} d t_2 \frac{1}{t_2-q_2^2}\int d^4k_3 d^4k_4\,\delta^4 \left(q_2-k_3-k_4 \right)\delta \left(k_3^2-m_Q^2 \right)\delta \left(k_4^2-m_Q^2 \right)  \nonumber \\
&& \frac{(-2\pi i)^2}{2\pi i} \int_{4m_Q^2}^{t_3^f} d t_3 \frac{1}{t_3-q_3^2}\int d^4k_5 d^4k_6\,\delta^4 \left(q_3-k_5-k_6 \right)\delta \left(k_5^2-m_Q^2 \right)\delta \left(k_6^2-m_Q^2 \right)  \nonumber \\
&&f_{\mu\nu}(k_1,k_2,k_3,k_4,k_5,k_6)\, ,
\end{eqnarray}
where the $f_{\mu\nu}(k_1,k_2,k_3,k_4,k_5,k_6)$ denotes  the numerator after carrying out the trace in the Dirac spinor space, the $t_1^f$, $t_2^f$ and $t_3^f$ are the upper bounds of the integrals. Then we accomplish  the integrals
$\int d^4k_1 d^4k_2$, $\int d^4k_3 d^4k_4$ and $\int d^4k_5 d^4k_6$ to obtain,
\begin{eqnarray}
\Pi_{\mu\nu}(p)&=&\int d^4 p_1 d^4 q_3\, \delta^4 \left(p-p_1-q_3 \right) \nonumber\\
&&\frac{(-2\pi i)^2}{2\pi i} \int_{16m_Q^2}^{r^f} d r \frac{1}{r-p_1^2}\int_{4m_Q^2}^{t_1^f} d t_1\int_{4m_Q^2}^{t_2^f} d t_2 \int_{4m_Q^2}^{t_3^f} d t_3
  \nonumber \\
&&\int d^4q_1 d^4q_2\,\delta^4 \left(p_1-q_1-q_2 \right)\delta \left(q_1^2-t_1 \right)\delta\left(q_2^2-t_2 \right) \frac{1}{t_3-q_3^2}\,\tilde{f}_{\mu\nu}(q_1,q_2,q_3,t_1,t_2,t_3) \nonumber \\
&=&\frac{(-2\pi i)^2}{2\pi i} \int_{36m_Q^2}^{\infty} d s \frac{1}{s-p^2} \int_{16m_Q^2}^{r^f} d r \int_{4m_Q^2}^{t_1^f} d t_1\int_{4m_Q^2}^{t_2^f} d t_2 \int_{4m_Q^2}^{t_3^f} d t_3 \nonumber\\
&&\int d^4 p_1 d^4 q_3\, \delta^4 \left(p-p_1-q_3 \right) \delta \left(p_1^2-r \right)\delta \left(q_3^2-t_3 \right) \nonumber \\
&& \frac{(-2\pi i)^2}{2\pi i}\int d^4q_1 d^4q_2\,\delta^4 \left(p_1-q_1-q_2 \right)\delta \left(q_1^2-t_1 \right)\delta \left(q_2^2-t_2 \right) \,\tilde{f}_{\mu\nu}(q_1,q_2,q_3,t_1,t_2,t_3) \, , \nonumber \\
\end{eqnarray}
where the $\tilde{f}_{\mu\nu}(q_1,q_2,q_3,t_1,t_2,t_3)$ denotes the lengthy expressions after accomplishing  the integrals $\int d^4k_1 d^4k_2$, $\int d^4k_3 d^4k_4$ and $\int d^4k_5 d^4k_6$, the $r^f$ is the upper bound of the integral. Finally, we accomplish the integrals $\int d^4 p_1 d^4 q_3$ and $\int d^4q_1 d^4q_2$ to obtain,
\begin{eqnarray}
\Pi_{\mu\nu}(p)&=& \int_{36m_Q^2}^{s_0}ds \int_{16m_Q^2}^{(\sqrt{s}-2m_Q)^2}dr \int_{4m_Q^2}^{(\sqrt{r}-2m_Q)^2}dt_1
\int_{4m_Q^2}^{(\sqrt{r}-\sqrt{t_1})^2}dt_2\int_{4m_Q^2}^{(\sqrt{s}-\sqrt{r})^2}dt_3\nonumber\\
&&\frac{1}{s-p^2}\left\{ \rho_{V}(s,r,t_1,t_2,t_3)\left(-g_{\mu\nu}+\frac{p_\mu p_\nu}{p^2} \right)+\rho_{S}(s,r,t_1,t_2,t_3)\frac{p_\mu p_\nu}{p^2}  \right\}+\cdots\, .\nonumber \\
\end{eqnarray}

Now we give an example to illustrate the endpoint divergence, at the lowest order, we often encounter the typical integral,
\begin{eqnarray}
I_{11}&=&\int d^4k_1d^4k_2
\frac{1}{k_1^2-m_1^2}\frac{1}{k_2^2-m_2^2}\delta^4\left( q-k_1-k_2\right)\, ,
\end{eqnarray}
and calculate it by using the Cutkosky's rules,
\begin{eqnarray}
I_{11}&=& \frac{(-2\pi i)^2}{2\pi
i}\int_{(m_1+m_2)^2}^{\infty}dt\frac{1}{t-q^2}\int d^4k_1d^4k_2
\delta^4(q-k_1-k_2)\delta(k_1^2-m_1^2)\delta(k_2^2-m_2^2)\nonumber\\
&=&\frac{(-2\pi i)^2}{2\pi
i}\int_{(m_1+m_2)^2}^{\infty}dt\frac{1}{t-q^2}\frac{\pi}{2}\frac{\sqrt{\lambda(t,m_1^2,m_2^2)}}{t}\,
,
\end{eqnarray}
which is free of end-point divergence. At the second Feynman
diagram in Fig.\ref{Feynman}, we often encounter the typical
integral,
\begin{eqnarray}
I_{22}&=& \int d^4k_1d^4k_2
\frac{1}{(k_1^2-m_1^2)^2}\frac{1}{(k_2^2-m_2^2)^2}\delta^4\left( q-k_1-k_2\right)\, ,
\end{eqnarray}
again we calculate it by using the Cutkosky's rules,
\begin{eqnarray}\label{I22AB}
I_{22}&=&\frac{\partial^2}{\partial A \partial B} \int d^4k_1d^4k_2
\frac{1}{k_1^2-A}\frac{1}{k_2^2-B}\delta^4\left( q-k_1-k_2\right)\mid_{A\to m_1^2; B\to m_2^2}\nonumber\\
&=&\frac{\partial^2}{\partial A \partial B} \frac{(-2\pi
i)^2}{2\pi
i}\int_{(\sqrt{A}+\sqrt{B})^2}^{\infty}dt\frac{1}{t-q^2}\int
d^4k_1d^4k_2\delta^4(q-k_1-k_2)\delta(k_1^2-A)\delta(k_2^2-B)\nonumber\\
&=&\frac{\partial^2}{\partial A \partial B}\frac{(-2\pi i)^2}{2\pi
i}\int_{(\sqrt{A}+\sqrt{B})^2}^{\infty}dt\frac{1}{t-q^2}\frac{\pi}{2}\frac{\sqrt{\lambda(t,A,B)}}{t}\nonumber\\
&=&\frac{(-2\pi i)^2}{2\pi
i}\int_{(m_1+m_2)^2}^{\infty}dt\frac{1}{t-q^2}\frac{\pi}{2}
\frac{2(m_1^2+m_2^2-t)}{\sqrt{\lambda(t,m_1^2,m_2^2)}^3}\, .
\end{eqnarray}
In the limit $m_1^2=m_2^2=m_Q^2$, we obtain
\begin{eqnarray} \label{ED-div-DR}
\int_{(m_1+m_2)^2}^{\infty}dt\frac{1}{t-q^2}\frac{1}{\sqrt{\lambda(t,m_1^2,m_2^2)}^3}&=&
\int_{4m_Q^2}^{\infty}dt\frac{1}{t-q^2}\frac{1}{\sqrt{t(t-4m_Q^2)}^3}\,
,
\end{eqnarray}
divergence at the end-point $t=4m_Q^2$. For the end-point divergence  $\frac{1}{\sqrt{t-4m_Q^2}^5}$ appears at the
first diagram in Fig.\ref{Feynman}, the calculations are analogous. The end-point divergences appear as a natural outcome of the calculations using the Cutkosky's rules, and are usually regularized by  adding a   mass term.

\section*{Acknowledgements}
This  work is supported by National Natural Science Foundation, Grant Number 12175068.

\end{document}